\documentclass[11pt]{article}

\usepackage{amsmath,amssymb}

\usepackage{mathptmx}
\newtheorem{teor}{Theorem}
\newtheorem{prop}{Proposition}

\newtheorem{definition}{Definition}
\newtheorem{remark}{Remark}
\def\beq{\begin{equation}}
\def\eeq{\end{equation}}
\def\bea{\begin{eqnarray}}
\def\eea{\end{eqnarray}}
\def\beann{\begin{eqnarray*}}
\def\eeann{\end{eqnarray*}}
\def\beasn{\begin{sneqnarray}}
\def\eeasn{\end{sneqnarray}}
\def\ben{\begin{enumerate}}
\def\een{\end{enumerate}}
\def\bit{\begin{itemize}}
\def\eit{\end{itemize}}
\def\proof{ ({\sl Proof\/}) }
\newcommand{\ds}{\displaystyle}
\def\derpar#1#2{\frac{\partial{#1}}{\partial{#2}}}

\def\qed{\ifvmode\Realemovelastskip\fi
{\unskip\nobreak\hfil\penalty50\hbox{}\nobreak\hfil \hbox{\vrule
height1.2ex width1.2ex}\parfillskip=0pt \finalhyphendemerits=0
\par\smallskip}}

\def\df{{\mit\Omega}}

\def\d{{\rm d}}

\def\Real{\mathbb{R}}

\def\Lie{\mathop{\rm L}\nolimits}
\def\inn{\mathop{i}\nolimits}
\def\Cinfty{{\rm C}^\infty}
\def\tabaddress#1{{\small\it\begin{tabular}[t]{c}#1
\\[1.2ex]\end{tabular}}}
\def\qed{\ifvmode\removelastskip\fi
{\unskip\nobreak\hfil\penalty50\hbox{}\nobreak\hfil \hbox{\vrule
height1.2ex width1.2ex}\parfillskip=0pt \finalhyphendemerits=0
\par\smallskip}}

\title{HIGHER-ORDER CARTAN SYMMETRIES IN $k$-SYMPLECTIC FIELD THEORY}
\author{\sc Narciso Rom\'an-Roy\thanks{{\bf e}-{\it mail}:
   nrr@ma4.upc.edu}
   \\
   \tabaddress{Departamento de Matem\'atica Aplicada IV.
   Edificio C-3, Campus Norte UPC\\
   C/ Jordi Girona 1. 08034 Barcelona. Spain}
   \\
{\sc Modesto Salgado\thanks{{\bf e}-{\it mail}:
modesto@zmat.usc.es},
      Silvia Vilari\~no\thanks{{\bf e}-{\it mail}:
svfernan@usc.es}} \\
  \tabaddress{Departamento de Xeometr\'ia e Topolox\'ia,
     Facultade de Matem\'{a}ticas. \\
     Universidade de Santiago de Compostela.
     15782-Santiago de Compostela. Spain}}

\begin{document}

\maketitle

\thispagestyle{empty}

   \bigskip   \bigskip

\begin{abstract}
 For $k$-symplectic Hamiltonian field theories,
we study  infinitesimal transformations generated by certain kinds of vector fields
which are not Noether symmetries, but which
allow us to obtain conservation laws by means of a suitable
generalization of the Noether theorem.
\end{abstract}




\vbox{\raggedleft AMS s.\,c.\,(2000): 70S05, 70S10, 53D05
  \\
PACS (1999):  11.10.Ef, 11.10.Kk, 02.40.Hw}\null

\markright{\sc N. Rom\'an-Roy {\it et al\/},
     \sl Higher-order Cartan symmetries in field theory}


\section{Introduction}

The $k$-symplectic formalism \cite{aw,mt2,fam} is the simplest generalization to field
theories of the standard symplectic formalism in autonomous Mechanics.
It allows usto give a geometric description of certain kinds of field theories: in
a local description, those theories whose Lagrangian
or Hamiltonian functions depend only on
the field coordinates and on the partial derivatives of the
fields, or on the corresponding moments, but not on the
base coordinates.
This formalism is based on the polysymplectic formalism
developed by G\"{u}nther \cite{gun}.

In a previous paper \cite{mod} we introduced the
notion of {\sl Cartan or Noether symmetry}, and
we stated Noether's theorem for Hamiltonian and Lagrangian systems
in $k$-symplectic field theories.
Noether's theorem associates conservation laws to Cartan or Noether
  symmetries. However, these kinds of symmetries do not exhaust the set of
  (general) symmetries. As is known, in mechanics there are
  dynamical symmetries which are not of Noether type, but
  which also generate conserved quantities
  (see \cite{LMR-99}, \cite{Ra-95}, \cite{Ra-97},
  for some examples). These are the so-called {\sl hidden
  symmetries}. Different attempts have been made
  to extend Noether's theorem in order
  to include these symmetries and the corresponding conserved
  quantities for mechanical systems (see for instance \cite{SC-81})
  and multisymplectic field theories (see \cite{EMR-99}).

  In this paper we present a generalization of the Noether
  theorem for $k$-symplectic Hamiltonian field theories, which is based in the approach of
  reference \cite{SC-81} for mechanical systems.
 This generalization allows us to obtain conservation laws associated to
 infinitesimal transformations generated by certain kinds of vector fields
 which are not Noether symmetries.

All manifolds
are real, paracompact, connected and $C^\infty$. All
maps are $C^\infty$. Sum over crossed repeated indices is understood.

\section{$k$-symplectic Hamiltonian systems}

(See \cite{fam}, \cite{rrsv}, \cite{mod} for details).

Let $(T^1_k)^*Q= T^*Q\oplus \stackrel{k}{\dots} \oplus T^*Q$
 be the \emph{bundle of $k^1$ covelocities} of an
$n$-dimensional differentiable manifold $Q$,
with projection  $\tau^*\colon (T^1_k)^*Q \to Q$. Natural
coordinates on $(T^1_k)^*Q$ are $(q^i , p^A_i);\, 1\leq
i\leq n,\, 1\leq A \leq k$.

The  \emph{ canonical $k$-symplectic structure} in $(T^1_k)^*Q$ is
 $(\omega^A,V)$,
where $V =\ker(\tau^*)_*$, and
$\omega^A = (\tau^*_A)^*\omega =-\d (\tau_A^*)^*\theta  =-\d\theta^A$;
$\omega=-\d\theta$ being the canonical symplectic structure in $T^*Q$
($\theta\in\df^1(T^*Q)$ is the \emph{ Liouville $1$-form\/}),
and $\tau^*_A\colon (T^1_k)^*Q \rightarrow T^*Q $ the
projection on the $A^{th}$-copy $T^*Q$ of $(T^1_k)^*Q$.
Locally
$$
\omega^A= -\d\theta^A=- \d(p^A_i \,\d q^i )=\d q^i\wedge\d p^A_{i}  \, .
$$

Given a diffeomorphism $\varphi\colon Q \to Q$,
its \emph{ canonical prolongation} to $(T^1_k)^*Q$  is the map
$(T^1_k)^*\varphi\colon(T^1_k)^*Q \to (T^1_k)^*Q$,
which is defined by
$$
(T^1_k)^*\varphi({\alpha_1}_q,\ldots,{\alpha_k}_q)=
(T^*\varphi({\alpha_1}_q),\ldots,T^* \varphi({\alpha_k}_q)) \ ,\
({\alpha_1}_q,\ldots,{\alpha_k}_q)\in (T^1_k)_q^*Q,\ q\in Q \, .
$$
If $Z\in{\mathfrak{X}}(Q)$ has $h_s\colon Q \to Q$ as  local $1$-parametric group;
 the \emph{ canonical lift} of $Z$ to $(T^1_k)_q^*Q$ is the vector field
$Z^{C*}\in{\mathfrak{X}}((T^1_k)^*Q)$ whose local $1$-parametric group is
 $(T^1_k)^*(h_s)\colon (T^1_k)^*Q\to(T^1_k)^*Q$.
Locally, if $\ds Z=Z^i\ds\frac{\partial}{\partial q^i}$ then
$\ds Z^{C*}= Z^i\frac{\partial}{\partial q^i} \, - \, p_j^A \ds
\frac{\partial Z^j} {\partial q^k} \frac{\partial}{\partial p_k^A}$.

\begin{definition}
 Let $T^1_kM=TM\oplus\stackrel{k}{\dots}\oplus TM$ be
the {\rm bundle of $k^1$ velocities} of a manifold~$M$. Let us denote by $\tau: T^1_kM \to M$ the canonical projection.
 \begin{itemize}
\item
A  {\rm $k$-vector field} on $M$ is a section
 ${\bf X} \colon M \longrightarrow T^1_kM$ of $\tau$.

A $k$-vector field ${\bf X}$ defines
a family of vector fields $X_{1}, \dots, X_{k}\in{\mathfrak{X}}(M)$ by
$X_A=\tau_A\circ{\bf X}$, where
$\tau_A\colon T^1_kM \rightarrow TM$ is the
projection on the $A^{th}$-copy $TM$ of $T^1_kM$.
\item
An {\rm integral section}  of
${\bf X}$ at a point $q\in M$, is a map
$\psi\colon U_0\subset \Real^k \rightarrow M$, with $0\in U_0$,
such that
$\psi(0)=q$, $\ds\psi_{*}(t)\left(\frac{\partial}{\partial t^A}\Big\vert_t\right)=X_{A}(\psi (t))$,
for every $t\in U_0$;
or what is equivalent,  $\psi$ satisfies that
${\bf X}\circ\psi=\psi^{(1)}$, where  $\psi^{(1)}$ is the first
prolongation of $\psi$  to $T^1_kM$ defined by
$$
\begin{array}{rccl}\label{1prolong}
\psi^{(1)}: & U_0\subset \Real^k & \longrightarrow & T^1_kM
\\\noalign{\medskip}
 & t & \longrightarrow & \psi^{(1)}(t)=
 \left(\psi_*(t)\left(\derpar{}{t^1}\Big\vert_t\right),\ldots,
\psi_*(t)\left(\derpar{}{t^k}\Big\vert_t\right)\right) \, .
 \end{array}
$$
 A $k$-vector field is
{\rm integrable} if there is an integral section at
every point of $M$.
\end{itemize}
\end{definition}

The set of $k$-vector fields on $M$ are denoted by ${\mathfrak{X}}^k(M)$.

Now take $M=(T^1_k)^*Q$. Let $H \colon (T^1_k)^*Q \to \Real$ be a \emph{ Hamiltonian function}.
 The family $((T^1_k)^*Q,\omega^A,H)$ is a \emph{ $k$-symplectic
Hamiltonian system}. The \emph{ Hamilton-de Donder-Weyl (HDW) equations}
associated to this system are
\begin{equation}
\label{he}
 \frac{\displaystyle
\partial H}{\displaystyle\partial q^i}\Big\vert_{\psi(t)}=
-\sum_{A=1}^k\frac{\displaystyle \partial\psi^A_i} {\displaystyle
\partial t^A}\Big\vert_t
\quad , \quad \frac{\displaystyle \partial H} {\displaystyle
\partial p^A_i}\Big\vert_{\psi(t)}= \frac{\displaystyle \partial\psi^i}{\displaystyle
\partial t^A}\Big\vert_t\, ,
\end{equation}
where $\psi\colon\Real^k\to (T^1_k)^*Q$,
$\psi(t)=(\psi^i(t),\psi^A_i(t))$, is a solution.

We denote by ${\mathfrak{X}}^k_H((T^1_k)^*Q)$ the set of $k$-vector fields
on $(T^1_k)^*Q$ solutions to
$$
\sum_{A=1}^k\inn(X_A)\omega^A=\d H\;.
 \label{generic}
$$
In a local system of canonical coordinates, each $X_A$ is locally given by
\ $\ds X_A =(X_A)^i\frac{\partial}{\partial q^i}+(X_A)_i^B\frac{\partial}{\partial p_i^B}$,\
and we obtain that the equation (\ref{generic}) is equivalent to the equations
 \begin{equation}
  \label{11}
 \frac{\partial H}{\partial q^i}=\, -\ds\sum_{A=1}^k\,(X_A)^A_i
 \quad  , \quad \frac{\partial H}{\partial p^A_i}=(X_A)^i\ .
\end{equation}

If ${\bf X}=(X_1,\dots,X_k)$ is an integrable $k$-vector field in $(T^1_k)^*Q$,
and $\psi\colon\Real^k\to (T^1_k)^*Q$ an integral section of
${\bf X}$, we have that $\psi(t)=(\psi^i(t),\psi^A_i(t))$ is a solution to the
HDW-equations (\ref{he}) if, and only if, ${\bf X}\in{\mathfrak{X}}^k_H((T^1_k)^*Q)$.
In fact, if $\psi(t)=(\psi^i(t),\psi^A_i(t))$ is an integral section of ${\bf X}$, then
 \begin{equation}
  \label{intsec}
\frac{\partial\psi^i}{\partial t^B}=(X_B)^i \quad , \quad
\frac{\partial\psi^A_i}{\partial t^B}=(X_B)^A_i \,  .
\end{equation}
and therefore (\ref{11}) are the HDW-equations (\ref{he}).

\begin{remark}
{\rm
We can define the vector bundle morphism
$$
\begin{array}{ccccc}
\omega^{\sharp} & \colon & T^1_k((T^1_k)^ *Q) & \to & T^*((T^1_k)^ *Q) \\
 & & (v_{p_1},,\ldots,v_{p_k}) & \mapsto & \displaystyle \sum_{A=1}^k\inn(v_{p_A})\omega^A_p
\end{array} \ ,
$$
and we denote with the same symbol its natural extension
$$
\begin{array}{ccccc}
\omega^{\sharp} & \colon & {\mathfrak{X}}^k((T^1_k)^ *Q)& \to & \df^1((T^1_k)^*Q) \\
 & & {\bf X}=(X_1,\dots,X_k) & \mapsto & \displaystyle \sum_{A=1}^k\inn(X_A)\omega^A
\end{array} \ .
$$
Then, the solutions to (\ref{generic}) are given by
${\bf X}+\ker\,\omega^{\sharp}$, where
${\bf X}$ is a particular solution.
}
\end{remark}

The equations (\ref{he}) and (\ref{generic}) are not equivalent because
not every solution to the HDW-equations
(\ref{he}) is an integral section of some integrable $k$-vector
field belonging to ${\mathfrak{X}}^k_H((T^1_k)^*Q)$, unless some additional conditions
are required. Thus, we assume the following condition
(which holds for a large class of mathematical applications and
physical field theories):

\begin{definition}
\label{hyp} A map $\psi\colon\Real^k\to (T^1_k)^*Q$, solution to the
equations (\ref{he}),
 is said to be an {\rm admissible solution} to the HDW-equations
for a $k$-symplectic Hamiltonian system $((T^1_k)^*Q,H)$, if ${\rm
Im}\,\psi$ is an embedded submanifold of $(T^1_k)^*Q$.

We say that $((T^1_k)^*Q,H)$  is an {\rm admissible $k$-symplectic
Hamiltonian system} when only admissible solutions to its HDW-equations are
considered.
\end{definition}

\begin{prop}\label{prop1}
Every admissible solution to the HDW-equations (\ref{he}) is an
integral section of an integrable $k$-vector field ${\bf
X}\in{\mathfrak{X}}^k_H((T^1_k)^*Q)$.
 \label{prop3}
\end{prop}
\proof\
 Let $\psi\colon\Real^k\to(T^1_k)^*Q$ be an admissible solution
to the HDW-equations (\ref{he}). By hypothesis, ${\rm Im}\,\psi$
is a $k$-dimensional submanifold of $(T^1_k)^*Q$. As $\psi$
is an embedding, we can define a $k$-vector field ${\bf
X}\vert_{{\rm Im}\,\psi}$ (at support on ${\rm Im}\,\psi$),
tangent to ${\rm Im}\,\psi$,~by
$$
 X_A(\psi(t))=(\psi)_*(t)\left(\frac{\partial}{\partial t^A}\Big\vert_t\right) \ ,
$$
which is a solution to (\ref{generic}) on the points of ${\rm
Im}\,\psi$, since (\ref{11}) holds on these points as a consequence
of (\ref{he}) and (\ref{intsec}).
 Furthermore, ${\rm Im}\,\psi$ is a
 submanifold of $(T^1_k)^*Q$; therefore we can extend  this
$k$-vector field ${\bf X}\vert_{{\rm Im}\,\psi}$ to an integrable
$k$-vector field  ${\bf X}\in{\mathfrak{X}}^k_H((T^1_k)^*Q)$ in such a way that
this extension is a solution to the equations (\ref{generic})
(note that these equations have solutions
everywhere on $(T^1_k)^*Q$), and which obviously
has $\psi$ as an integral section. This extension is made at least
locally, and then the global $k$-vector field is constructed using
partitions of unity.
 \qed

In this way, for admissible $k$-symplectic Hamiltonian systems,
 the field equations (\ref{generic}) are a geometric
version of the HDW-equations (\ref{he}).

\section{Symmetries and conservation laws}
\protect\label{scl}

\begin{definition}
{\rm (Olver \cite{olver})}
A {\rm conservation law} or a {\rm conserved quantity} of a
$k$-symplectic Hamiltonian system $((T^1_k)^*Q,\omega^A,H)$ is a
map ${\cal F}=({\cal F}_1,\ldots,{\cal F}_k)\colon (T^1_k)^*Q\to
\Real^k$ such that the
divergence of ${\cal F}\circ\psi=({\cal F}_1\circ\psi,\ldots,{\cal
F}_k\circ\psi)\colon\Real^k\to\Real^k$ is zero for every
$\psi\colon\Real^k\to (T^1_k)^*Q$ solution to the Hamilton-de
Donder-Weyl equations (\ref{he}); that is,
$\ds \sum_{A=1}^k\frac{\partial({\cal F}_A\circ\psi)}{\partial t^A}=0$.
\label{olver}
\end{definition}

For admissible $k$-symplectic Hamiltonian systems,
conserved quantities can be characterized as follows:

  \begin{prop}Let $((T^1_k)^*Q,H)$  be an  admissible $k$-symplectic
Hamiltonian system.

A map ${\cal F}=({\cal F}_1 , \ldots , {\cal F}_k)\colon
(T^1_k)^*Q\to \Real^k$ is a conservation law of an admissible
$k$-symplectic Hamiltonian system if, and only if,
 for every integrable $k$-vector field
${\bf X}=(X_1,\dots,X_k)\in{\mathfrak{X}}^k_H((T^1_k)^*Q)$, we have that
$\ds \sum_{A=1}^k\Lie({X_A}){\cal F}_A=0$.
\label{cc}
  \end{prop}
\proof\
Let  ${\cal F}=({\cal F}^1,\ldots,{\cal F}^k)$ be a conservation law
and ${\bf X}=(X_1,\dots,X_k)\in{\mathfrak{X}}^k_H((T^1_k)^*Q)$ an
integrable $k$-vector field. If $\psi\colon \Real^k\to
(T^1_k)^* Q$ is an integral section of ${\bf X}$ then:
\ben
\item  We have that $\psi$ is a solution
to the Hamilton-de Donder-Weyl equation (\ref{he}).
\item By definition of integral section, we have
$ X_A(\psi(t))=\psi_*(t)\left(\derpar{}{t^A}\Big\vert_t\right) $ .
\end{enumerate}
Therefore
  \[\sum_{A=1}^k\Lie(X_A){\cal F}^A=
\sum_{A=1}^k\psi_*(t)\left(\derpar{}{t^A}\Big\vert_t\right)({\cal
F}^A)=\sum_{A=1}^k\derpar{({\cal F}^A\circ\psi)}{t^A}\Big\vert_t = 0 \, \,  .\]

Conversely, let us suppose that  every integrable $k$-vector field
${\bf X}=(X_1,\dots,X_k)$ in ${\mathfrak{X}}^k_H((T^1_k)^*Q)$
satisfies $\ds \sum_{A=1}^k\Lie({X_A}){\cal F}_A=0$,
and let  $\psi\colon \Real^k\to  (T^1_k)^* Q$ be an admissible solution to the
HDW-equations (\ref{he}). By Proposition
\ref{prop1} there exists a $k$-vector field ${\bf X}\in{\mathfrak{X}}^k_H((T^1_k)^*Q)$
such that 
$$
 X_A(\psi(t))=(\psi)_*(t)\left(\frac{\partial}{\partial
t^A}\Big\vert_t\right)$$ 
 Thus, since  $\ds \sum_{A=1}^k\Lie(X_A){\cal F}^A= 0$, from the above identity we obtain that 
  $$\sum_{A=1}^k\derpar{({\cal F}^A\circ\psi)}{t^A}\Big\vert_t=0\, \, . $$
\qed

\begin{definition}
 \label{symH}
Let $((T^1_k)^*Q,\omega^A,H)$ be a $k$-symplectic Hamiltonian system.
\ben
\item
A {\rm symmetry} is a diffeomorphism
$\Phi\colon (T^1_k)^*Q \to (T^1_k)^*Q$ such that for
every solution $\psi$ to  the HDW equations (\ref{he}), we
have that $\Phi\circ\psi$ is also a solution.

If $\Phi=(T^1_k)^*\varphi$ for some
$\varphi\colon Q\to Q$, the symmetry
$\Phi$ is said to be {\rm natural}.
\item
 An {\rm infinitesimal symmetry}
 is a vector field $Y\in{\mathfrak{X}}((T^1_k)^*Q)$ whose local flows are local symmetries.

If $Y=Z^{C*}$ for
$Z\in{\mathfrak{X}} (Q)$, the infinitesimal symmetry
$Y$ is said to be {\rm natural}.
\een
\end{definition}

\begin{prop}
Let $((T^1_k)^*Q,\omega^A,H)$ be an admissible $k$-symplectic Hamiltonian system.
A diffeomorphism $\Phi\colon (T^1_k)^*Q\to (T^1_k)^*Q$ is
a symmetry if, and only if, for every integrable $k$-vector field
${\bf X}=(X_1,\dots,X_k)\in{\mathfrak{X}}^k_H((T^1_k)^*Q)$
we have that
 $\Phi_*{\bf X}=(\Phi_*X_1,\dots,\Phi_*X_k)\in{\mathfrak{X}}^k_H((T^1_k)^*Q)$,
it is integrable, and its integral sections are $\Phi\circ\psi$, for any integral section  $\psi$
of ${\bf X}$.
\label{pro4}
\end{prop}
\proof\
Let $\Phi\colon (T^1_k)^*Q\to (T^1_k)^*Q$ be a diffeomorphism and
${\bf X}=(X_1,\dots,X_k)$ an integrable $k$-vector field in ${\mathfrak{X}}^k_H((T^1_k)^*Q)$.
Then every integral section $\psi$ of ${\bf X}$ is a solution to theHDW-equations (\ref{he})
and satisfy $X_A(\psi(t))=\psi_*(t)(\partial/\partial t^A)$.
From this we have $\Phi_*(\psi(t))X_A(\psi(t))=(\Phi\circ \psi)_*(\partial / \partial t^A)$,
 thus $\Phi\circ \psi $
is an integral section of $\Phi_* {\bf X}$, and so $\Phi_* {\bf X}$ is integrable.

    Now, since $\Phi$ is a symmetry, then $\Phi\circ \psi  $ is a
solution to theHDW-equations (\ref{he})  and
as it is an integral section of $\Phi_* {\bf X}$ we deduce that
$\Phi_* {\bf X}\in{\mathfrak{X}}^k_H((T^1_k)^*Q)$.

Let $\psi$ be an admisible solution to the HDW-equations (\ref{he}), then by Proposition \ref{prop1},
 there exists ${\bf X}\in{\mathfrak{X}}^k_H((T^1_k)^*Q)$
 such that $\psi$ is an integral section of ${\bf X}$. Then $\Phi\circ \psi$
 is an integral section of $\Phi_*{\bf X}\in{\mathfrak{X}}^k_H((T^1_k)^*Q)$,
 and thus $\Phi\circ \psi$ is a solution to the HDW-equations (\ref{he}) .
\qed

As a consequence of this,  if $\Phi$ is a symmetry we have that
$\Phi_*{\bf X}-{\bf X}\in\ker\,\omega^{\sharp}$.

\begin{prop}
Let $((T^1_k)^*Q,\omega^A,H)$ be an admissible $k$-symplectic Hamiltonian system.
If $Y\in{\mathfrak{X}}((T^1_k)^*Q)$ is an
infinitesimal symmetry, then for every integrable $k$-vector field
${\bf X}=(X_1,\dots,X_k)\in{\mathfrak{X}}^k_H((T^1_k)^*Q)$
we have that $[Y,{\bf X}]=([Y,X_1],\dots,[Y,X_k])\in\ker\,\omega^{\sharp}$.
\label{pro5}
\end{prop}
\proof\
As $Y$ is an infinitesimal symmetry,
denoting by $F_t$ the local $1$-parameter groups of diffeomorphisms
generated by $Y$, we have that
$F_{t*}{\bf X}-{\bf X}\in\ker\,\omega^{\sharp}$. Then,
if $\{{\bf Z}^1,\ldots,{\bf Z}^r\}=\{(Z^1_1,\ldots,Z^1_k),\ldots,(Z^r_1,\ldots,Z^r_k)\}$ is
a local basis of $\ker\,\omega^{\sharp}$, we have that
$F_{t*}{\bf X}-{\bf X}=g_\alpha {\bf Z}^\alpha$, $\alpha=1,\ldots,r$,
with $g_\alpha\colon\Real\times(T^1_k)^*Q\to\Real$
(they are functions that depend on $t$); that is
$$
F_{t*}{\bf X}-{\bf X}=(F_{t*}X_1-X_1,\ldots,F_{t*}X_k-X_k)=
(g_\alpha Z_1^\alpha,\ldots,g_\alpha Z_k^\alpha)=
g_\alpha {\bf Z}^\alpha \ .
$$
 Therefore
\beann
[Y,{\bf X}]&=&\Lie(Y){\bf X}=(\Lie(Y)X_1,\ldots\Lie(Y)X_k)=
\left( \lim_{t\to 0}\frac{F_{t*}X_1-X_1}{t},\ldots, \lim_{t\to 0}\frac{F_{t*}X_k-X_k}{t} \right) \\
&=&
\left( \lim_{t\to 0}\frac{g_\alpha}{t}Z_1^\alpha,\ldots,
 \lim_{t\to 0}\frac{g_\alpha}{t}Z_k^\alpha \right)=
(f_\alpha Z_1^\alpha,\ldots,f_\alpha Z_k^\alpha)=
f_\alpha {\bf Z}^\alpha\in\ker\,\omega^{\sharp} \ ,
\eeann
where $f_\alpha\colon(T^1_k)^*Q\to\Real$.
\qed

\begin{prop}
Let $((T^1_k)^*Q,\omega^A,H)$ be an admissible $k$-symplectic Hamiltonian system.
If $Y\in{\mathfrak{X}}((T^1_k)^*Q)$ is an
infinitesimal symmetry, then for every ${\bf Z}\in\ker\,\omega^{\sharp}$,
we have that $[Y,{\bf Z}]\in\ker\,\omega^{\sharp}$.
\label{pro6}
\end{prop}
\proof\
For every ${\bf Z}\in\ker\,\omega^{\sharp}$,
there exist integrable $k$-vector fields
${\bf X},{\bf X'}\in{\mathfrak{X}}^k_H((T^1_k)^*Q)$
such that ${\bf X'}-{\bf X}={\bf Z}$; therefore
$[Y,{\bf Z}]=[Y,{\bf X'}]-[Y,{\bf X}]\in\ker\,\omega^{\sharp}$,
since $[Y,{\bf X'}],[Y,{\bf X}]\in\ker\,\omega^{\sharp}$,
by Proposition \ref{pro5}.
\qed

\section{Higher-order Cartan symmetries. Noether's theorem}
\protect\label{hocs}

Noether's theorem allows us to associate conservation laws to certain kinds of symmetries:
the so-called {\sl infinitesimal Cartan} or {\sl Noether symmetries}, which are
 vector fields $Y\in{\mathfrak{X}}((T^1_k)^*Q)$  such that: \
{\rm (i)}  $\Lie(Y)\omega^A=0$, \ and  \
{\rm (ii)} $\Lie(Y)H=0$ (see \cite{mod}).

Now we introduce new kinds of generators of conservation laws
which are not of this type
(we restrict ourselves to the infinitesimal case).

  \begin{definition}
Let $((T^1_k)^*Q,\omega^A,H)$ be a $k$-symplectic Hamiltonian system.
   A vector field $Y\in{\mathfrak{X}}((T^1_k)^*Q)$ is said to be
an {\rm infinitesimal Cartan} or {\rm  Noether symmetry of order $n$} if:
  \ben
\item
$Y$ is an  infinitesimal symmetry.
  \item
  $\Lie^n(Y)\omega^A:=\overbrace{\Lie(Y)\ldots\Lie(Y)}^{n}\omega^A=0$, but
  $\Lie^m(Y)\omega^A\not= 0$, for $m<n$.
\item
 $\Lie(Y)H=0$.
  \een
In the particular case that $Y=Z^{C*}$ for some $Z\in{\mathfrak{X}} (Q)$,
the infinitesimal Cartan (Noether) symmetry of order $n$ is said to be {\rm natural}.
  \label{CNsymn}
  \end{definition}

  For $n=1$ we recover the definition of  infinitesimal  Cartan (Noether) symmetry.
  Observe that infinitesimal Cartan symmetries of order $n>1$ are not
infinitesimal  Cartan symmetries.

  \begin{prop}
  If $Y\in{\mathfrak{X}} ((T^1_k)^*Q)$ is a  infinitesimal Cartan symmetry of order $n$
  of a $k$-symplectic Hamiltonian system, then the forms
  $\Lie^{n-1}(Y)\inn(Y)\omega^A\in\df^1((T^1_k)^*Q)$ are closed.
\label{p3}
  \end{prop}
  \proof\
  From the definition \ref{CNsymn}, we obtain
  $$
  0=\Lie^n(Y)\omega^A=\Lie^{n-1}(Y)\Lie(Y)\omega^A=
  \Lie^{n-1}(Y)\d\inn(Y)\omega^A=\d\Lie^{n-1}(Y)\inn(Y)\omega^A \ .
  $$
  \qed

\begin{prop}
Let $Y\in{\mathfrak{X}}((T^1_k)^*Q)$ be an  infinitesimal Cartan symmetry of order $n$
of a $k$-symplectic Hamiltonian system $((T^1_k)^*Q,\omega^A,H)$.
Then, for every $p\in (T^1_k)^*Q$, there
is an open neighbourhood $U_p\ni p$, such that:
 \ben
 \item
  There exist $g^A\in\Cinfty(U_p)$, which are unique up to constant
 functions, such that
\beq
 \Lie^{n-1}(Y)\inn(Y)\omega^A=\d g^A, \qquad \mbox{\rm (on
$U_p$)} \ .
 \label{bis}
 \eeq
 \item
 There exist $\xi^A\in\Cinfty(U_p)$,
verifying that $\Lie^n(Y)\theta^A=\d\xi^A$, on $U_p$; and then
\beq
 g^A=\inn(Y)\theta^A-\xi^A, \qquad \mbox{\rm (up to a constant
function, on $U_p$)}
 \label{bisdos}
 \eeq
 \een
 \label{structurebis}
\end{prop}
\proof\
 \ben
 \item
  It is an immediate consequence of Proposition \ref{p3} and the Poincar\'e Lemma.
  \item
We have that
$$
\d\Lie^n(Y)\theta^A=\Lie^n(Y)\d\theta^A=-\Lie^n(Y)\omega^A=0
$$
and hence $\Lie^n(Y)\theta^A$ are closed forms. Therefore, by the
Poincar\'e Lemma, there exist $\xi^A\in\Cinfty(U_p)$, verifying
that $\Lie^n(Y)\theta^A=\d\xi^A$, on $U_p$. Furthermore, as
(\ref{bis}) holds in $U_p$, we obtain that \beann \d\xi^A
&=&\Lie^n(Y)\theta^A = \Lie^{n-1}(Y)\Lie(Y)\theta^A=
\Lie^{n-1}(Y)\{\d\inn(Y)\theta^A+\inn(Y)\d\theta^A\}
\\ &=&
\d\Lie^{n-1}(Y)\inn(Y)\theta^A-\Lie^{n-1}(Y)\inn(Y)\omega^A=\d
\{\inn(Y)\theta^A-g^A\}
  \eeann
and thus (\ref{bisdos}) holds.
 \qed
 \een

  Finally, Noether's theorem
  can be generalized for these higher-order  Cartan symmetries
and admissible  $k$-symplectic Hamiltonian systems as follows:

  \begin{teor}
  {\rm (Noether):}
  If $Y\in{\mathfrak{X}} ((T^1_k)^*Q)$ is an infinitesimal  Cartan symmetry of order $n$
  of an admissible $k$-symplectic Hamiltonian system $((T^1_k)^*Q,\omega^A,H)$, then
$$
  g=(g^1,\ldots,g^k)=(\inn(Y)\theta^1-\xi^1,\ldots,\inn(Y)\theta^k-\xi^k)
$$
is a conserved quantity; that is, for every integrable $k$-vector field
 ${\bf X}=(X_1,\ldots,X_k)\in{\mathfrak{X}}^k_H((T^1_k)^*Q)$,
  we have that \
  $\ds \sum_{A=1}^k\Lie(X_A)g^A=0$ \ (on $U_p$).
  \label{Nthgen}
  \end{teor}
  \proof\
  If ${\bf X}=(X_1,\ldots ,X_k)\in{\mathfrak{X}}^k_H((T^1_k)^*Q)$,
  taking (\ref{bis}) into account we have
$$
  \sum_{A=1}^k\Lie(X_A)g^A =
  \sum_{A=1}^k\inn(X_A)\d g^A=
  \sum_{A=1}^k\inn(X_A)\Lie^{n-1}(Y)\inn(Y)\omega^A \ .
$$
Then, if $n=2$, we have \beann
  \sum_{A=1}^k\Lie(X_A)g^A &=&
  \sum_{A=1}^k\inn(X_A)\Lie(Y)\inn(Y)\omega^A=
  \sum_{A=1}^k\{\Lie(Y)\inn(X_A)-\inn([Y,X_A])\}\inn(Y)\omega^A
\\ &=&
  \sum_{A=1}^k\{-\Lie(Y)\inn(Y)\inn(X_A)+\inn(Y)\inn([Y,X_A])\}\omega^A
\\ &=&
  -\Lie(Y)\inn(Y)\d H=-\Lie^2(Y)H=0 \ ,
\eeann
 since as $Y$ is an infinitesimal Cartan symmetry of order $n$,
it is a symmetry; then $\Lie(Y) H=0$ and, by Proposition \ref{pro5},
$[Y,{\bf X}]\in\ker\,\omega^{\sharp}$, and hence
$\ds\sum_{A=1}^k\inn([Y,X_A])\omega^A=0$.

If $n=3$, by an analogous reasoning, we obtain
  \beann
  \sum_{A=1}^k\Lie(X_A)g^A &=&
  \sum_{A=1}^k\inn(X_A)\Lie^2(Y)\inn(Y)\omega^A=
  \sum_{A=1}^k\inn(X_A)\Lie(Y)\Lie(Y)\inn(Y)\omega^A
  \\ &=&
  \sum_{A=1}^k\{\Lie(Y)\inn(X_A)-\inn([Y,X_A])\}\Lie(Y)\inn(Y)\omega^A
  \\ &=&
  \sum_{A=1}^k\{\Lie(Y)\inn(X_A)\Lie(Y)-\inn([Y,X_A])\Lie(Y)\}\inn(Y)\omega^A
\\ &=&
  \sum_{A=1}^k\{\Lie^2(Y)\inn(X_A)-2\Lie(Y)\inn([Y,X_A])+\inn([Y,[Y,X_A]])\}\inn(Y)\omega^A \\
&=&
\sum_{A=1}^k\Lie^2(Y)\inn(X_A)\omega^A=\Lie^2(Y)H=0 \ ,
   \eeann
since, by Proposition \ref{pro5},
$[Y,{\bf X}]\in\ker\,\omega^{\sharp}$ and,
by Proposition~\ref{pro6}, $\ds\sum_{A=1}^k\inn([Y,[Y,X_A]])\omega^A=~0$.

For $n>3$, we arrive at the same result by
  repeating the above procedure $n-2$ times.
Thus, taking into account Proposition \ref{cc},
we have proved that $g=(g^1,\ldots,g^k)$ is a conservation law.
  \qed

{\bf Remark}:
{\sl $k$-symplectic Lagrangian systems} can be defined in
 $T^1_kQ= TQ\oplus \stackrel{k}{\dots} \oplus TQ$,
starting from a Lagrangian function $L\in\Cinfty(T^1_kQ)$, and
using the canonical structures of this $k$-tangent bundle
for defining a family of $k$ {\sl Lagrangian forms}
$\omega_L^A\in\df^2(T^1_kQ)$, and the {\sl Energy Lagrangian function}
$E_L\in\Cinfty(T^1_kQ)$ (see \cite{fam}, \cite{mod}). Then,
if the Lagrangian is regular, $(T^1_kQ,\omega_L^A, E_L)$ is
a $k$-symplectic Hamiltonian system with Hamiltonian function $E_L$, and
all the definitions and results in Sections \ref{scl} and \ref{hocs}
are applied to this case.

\subsection*{Acknowledgments}
We acknowledge the financial support of the
\emph{ Ministerio de Educaci\'on y Ciencia},
projects
MTM2006-27467-E and MTM2005-04947.
We thank Mr. Jeff Palmer for his assistance in preparing the
English version of the manuscript.

\end{document}